\documentclass[a4paper,11pt]{article}
\usepackage{jinstpub} 
\usepackage{subcaption}

\title{\boldmath The acrylic vessel for JSNS$^{2}$-II neutrino target}

\author[a]{C.D. Shin,}
\author[b]{S. Ajimura,}
\author[c]{M. K. Cheoun,}
\author[d]{J. H. Choi,}
\author[e]{J. Y. Choi,}
\author[f,i]{T. Dodo,}
\author[g]{J. Goh,}
\author[h]{\\K. Haga,}
\author[h]{M. Harada,}
\author[i,h]{S. Hasegawa,}
\author[b]{T. Hiraiwa,}
\author[g]{W. Hwang,}
\author[j]{T. Iida,}
\author[e]{ H. I. Jang,}
\author[k]{\\J. S. Jang,}
\author[l]{H. Jeon,}
\author[l]{S. Jeon,}
\author[m]{K. K. Joo,}
\author[l]{D. E. Jung,}
\author[n]{S. K. Kang,}
\author[h]{Y. Kasugai,}
\author[o]{\\T. Kawasaki,}
\author[p]{E. J. Kim,}
\author[m]{J. Y. Kim,}
\author[q]{S. B. Kim,}
\author[r]{W. Kim,}
\author[h]{H. Kinoshita,}
\author[o]{T. Konno,}
\author[a]{\\D. H. Lee,}
\author[m]{I. T. Lim,}
\author[s]{C. Little,}
\author[s]{E. Marzec,}
\author[a]{T. Maruyama, \footnote{Corresponding author.}}
\author[h]{S. Masuda,}
\author[h]{S. Meigo,}
\author[a]{\\S. Monjushiro,}
\author[m]{D. H. Moon,}
\author[b]{T. Nakano,}
\author[t]{M. Niiyama,}
\author[a]{K. Nishikawa,}
\author[d]{M. Y. Pac,}
\author[m]{\\H. W. Park,}
\author[r]{J. S. Park,}
\author[m]{R. G. Park,}
\author[u]{S. J. M. Peeters,}
\author[v,l]{C. Rott,}
\author[h]{K. Sakai,}
\author[h]{\\S. Sakamoto,}
\author[b]{T. Shima,}
\author[s]{J. Spitz,}
\author[f]{F. Suekane,}
\author[b]{Y. Sugaya,}
\author[h]{K. Suzuya,}
\author[a]{M. Taira,}
\author[j]{\\Y. Takeuchi,}
\author[h]{Y. Yamaguchi,}
\author[w]{M. Yeh,}
\author[d]{I. S. Yeo,}
\author[g]{C. Yoo}
\author[l]{and I. Yu}

\affiliation[a]{High Energy Accelerator Research Organization (KEK),\\ 1-1 Oho, Tsukuba, Ibaraki, 305-0801, Japan}
\affiliation[b]{Research Center for Nuclear Physics, Osaka University,\\ 10-1 Mihogaoka, Ibaraki, Osaka, 567-0047, Japan}
\affiliation[c]{Department of Physics, Soongsil University,\\ 369 Sangdo-ro, Dongjak-gu, Seoul, 06978, Korea}
\affiliation[d]{Laboratory for High Energy Physics, Dongshin University,\\ 67, Dongshindae-gil, Naju-si, Jeollanam-do, 58245, Korea}
\affiliation[e]{Department of Fire Safety, Seoyeong University,\\ 1 Seogang-ro, Buk-gu, Gwangju, 61268, Korea}
\affiliation[f]{Research Center for Neutrino Science, Tohoku University,\\ 6-3 Azaaoba, Aramaki, Aoba-ku, Sendai 980-8578, Japan}
\affiliation[g]{Department of Physics, Kyung Hee University,\\ 26, Kyungheedae-ro, Dongdaemun-gu, Seoul 02447, Korea}
\affiliation[h]{J-PARC Center, JAEA,\\ 2-4 Shirakata, Tokai-mura, Naka-gun, Ibaraki 319-1195, Japan}
\affiliation[i]{Advanced Science Research Center, JAEA,\\ 2-4 Shirakata, Tokai-mura, Naka-gun, Ibaraki 319-1195, Japan}
\affiliation[j]{Faculty of Pure and Applied Sciences, University of Tsukuba,\\ Tennodai 1-1-1, Tsukuba, Ibaraki, 305-8571, Japan}
\affiliation[k]{GIST College, Gwangju Institute of Science and Technology,\\ 123 Cheomdangwagi-ro, Buk-gu, Gwangju, 61005, Korea}
\affiliation[l]{Department of Physics, Sungkyunkwan University,\\ 2066, Seobu-ro, Jangan-gu, Suwon-si, Gyeonggi-do, 16419, Korea}
\affiliation[m]{Department of Physics, Chonnam National University,\\ 77, Yongbong-ro, Buk-gu, Gwangju, 61186, Korea}
\affiliation[n]{School of Liberal Arts, Seoul National University of Science and Technology,\\ 232 Gongneung-ro, Nowon-gu, Seoul, 139-743, Korea}
\affiliation[o]{Department of Physics, Kitasato University,\\ 1 Chome-15-1 Kitazato, Minami Ward, Sagamihara, Kanagawa, 252-0329, Japan}
\affiliation[p]{ Division of Science Education, Chonbuk National University,\\ 567 Baekje-daero, Deokjin-gu, Jeonju-si, Jeollabuk-do, 54896, Korea}
\affiliation[q]{School of Physics, Sun Yat-sen (Zhongshan) University,\\ Haizhu District, Guangzhou, 510275, China}
\affiliation[r]{Department of Physics, Kyungpook National University,\\ 80 Daehak-ro, Buk-gu, Daegu, 41566, Korea}
\affiliation[s]{University of Michigan,\\ 500 S. State Street, Ann Arbor, MI 48109, U.S.A.}
\affiliation[t]{Department of Physics, Kyoto Sangyo University,\\ Motoyama, Kamigamo, Kita-Ku, Kyoto-City, 603-8555, Japan}
\affiliation[u]{Department of Physics and Astronomy, University of Sussex,\\ Falmer, Brighton, BN1 9RH, U.K.}
\affiliation[v]{Department of Physics and Astronomy, University of Utah,\\ 201 Presidents' Cir, Salt Lake City, UT 84112, U.S.A}
\affiliation[w]{Brookhaven National Laboratory,\\ Upton, NY 11973-5000, U.S.A.}

\emailAdd{takasumi.maruyama@kek.jp}

\abstract
{
	The JSNS$^{2}$ (J-PARC Sterile Neutrino Search at J-PARC Spallation Neutron Source) is an experiment designed for the search for sterile neutrinos. The experiment is currently at the stage of the second phase named JSNS$^{2}$-II with two detectors at near and far locations from the neutrino source. One of the key components of the experiment is an acrylic vessel, that is used for the target volume for the detection of the anti-neutrinos. The specifications, design, and measured properties of the acrylic vessel are described.
}


\begin{document}
	\maketitle
	\flushbottom
	
	\section{Introduction}
	\label{sec:intro}
	
	The JSNS$^{2}$ (J-PARC Sterile Neutrino Search at J-PARC Spallation Neutron Source) experiment~\cite{JSNS2_proposal, JSNS2_TDR} began physics data taking, following a short commissioning phase, from 2020. The main scientific objective is the search for sterile neutrinos.
 The JSNS$^{2}$ detector is located at the Material and Life science experimental Facility (MLF) in J-PARC. 3 GeV protons contained within precisely timed bunches of the MLF proton beam with a designed beam power of 1 MW, impede on MLF's liquid mercury target to generate neutrinos via muon decay-at-rest ($\mu$DAR). The neutrino oscillation ($\bar{\nu}_{\mu} \to \bar{\nu}_{e}$) can be observed via inverse beta decay (IBD) reaction in the detector if sterile neutrinos exist. Since the neutrino source ($\mu$DAR) and detection principle (IBD) are the same as the LSND experiment~\cite{cite:LSND}, the JSNS$^{2}$ experiment can provide a direct test of the observed anomaly in the LSND experiment.
	
	\begin{figure}[htbp]
		\centering
		\includegraphics[width=.58\textwidth]{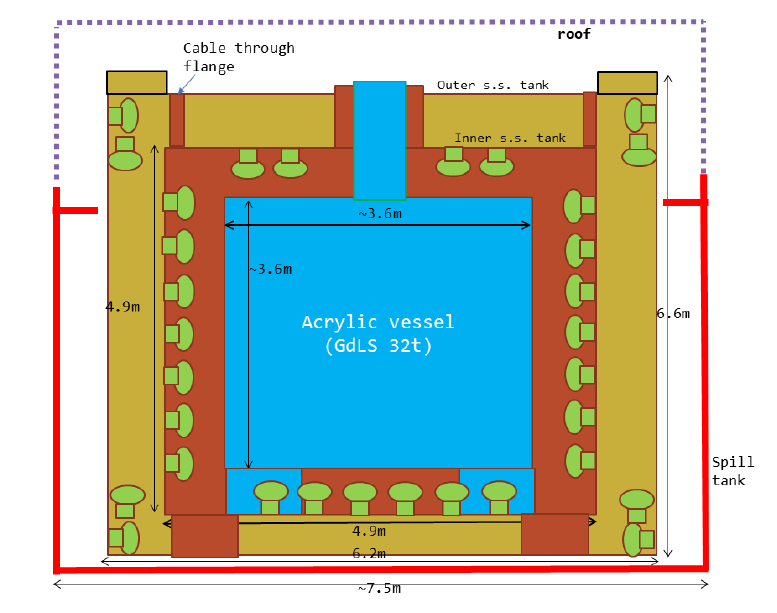}
		\caption{Structure of the JSNS$^{2}$-II detector}
		\label{fig:Detector}
	\end{figure}
	
	JSNS$^{2}$-II is the second phase of the JSNS$^{2}$ experiment with two detectors \cite{JSNS2_II_TDR}. The existing JSNS$^{2}$ detector will be utilized as the near detector with a 24 m baseline, and the newly constructed second detector will serve as a far detector, located at 48 m from the neutrino source. Figure \ref{fig:Detector} shows a schematic view of the second detector, which will be located outside next to the MLF building. The conceptional design of the far detector is almost identical to the JSNS$^{2}$ detector. The detector consists of three layers, which consist of a target, a catcher, and a veto region. The far detector target will be filled with 32 tonnes of Diisopropylnaphthalene (DIN) dissolved Gd-loaded liquid scintillator (Gd-LS). The amount of DIN is 10\% of Gd-LS, providing improved pulse shape discrimination (PSD) performance. Other regions such as a catcher and a veto will have 131 tons of the pure liquid scintillator (pure LS). Since different liquids are used for the target and the catcher regions, the detector target region must be separated physically, and transparent to allow photons to pass through. Similar to successful designs in reactor neutrino experiments \cite{DC_detector, RENO_acrylic, DayaBay_acrylic},  an acrylic vessel is chosen to separate the JSNS$^{2}$ target region.

	\section{Design of the acrylic vessel}
 
	The JSNS$^{2}$-II acrylic vessel is made of Polymethylmethacrylate (($\textrm{C}_{5}\textrm{H}_{8}\textrm{O}_{2}$)$_{\textrm{n}}$) called PMMA. This material is widely used due to its desirable optical properties. Especially, PMMA has good transparency in the region of scintillation emission and a similar refractive index to the LS. This reduces the reflection of light at the acrylic vessel due to a difference in refractive indices. In addition, an acrylic is an organic compound, thus the concentration of the radioactive isotopes are small \cite{DayaBay_acrylic}. 
	
	\begin{figure}[htbp]
		\centering
		\includegraphics[width=.61\textwidth]{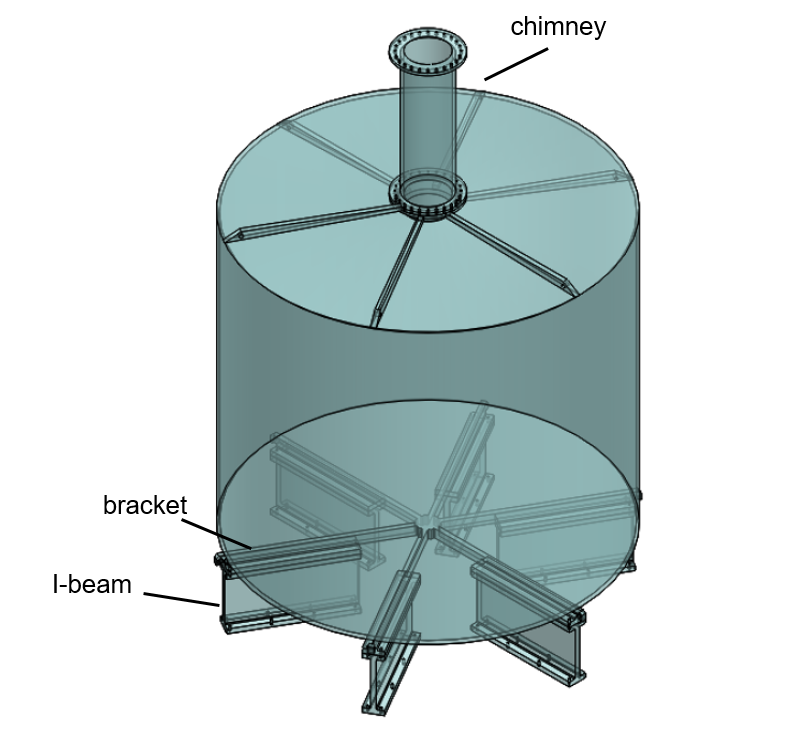}
		\caption{The design of the acrylic vessel.}
		\label{fig:Acrylic_vessel}
	\end{figure}
	
\begin{figure}[htbp]
		\centering
		\begin{subfigure}[b]{0.39\textwidth}
			\includegraphics[width=.75\textwidth]{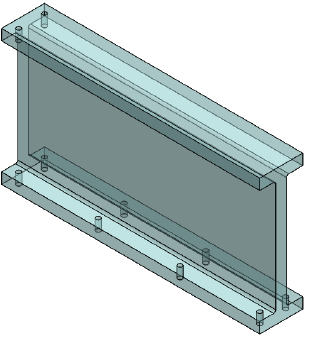}
			\caption{}
			\qquad
		\end{subfigure}
  \\
		\begin{subfigure}[b]{0.59\textwidth}
			\includegraphics[width=.98\textwidth]{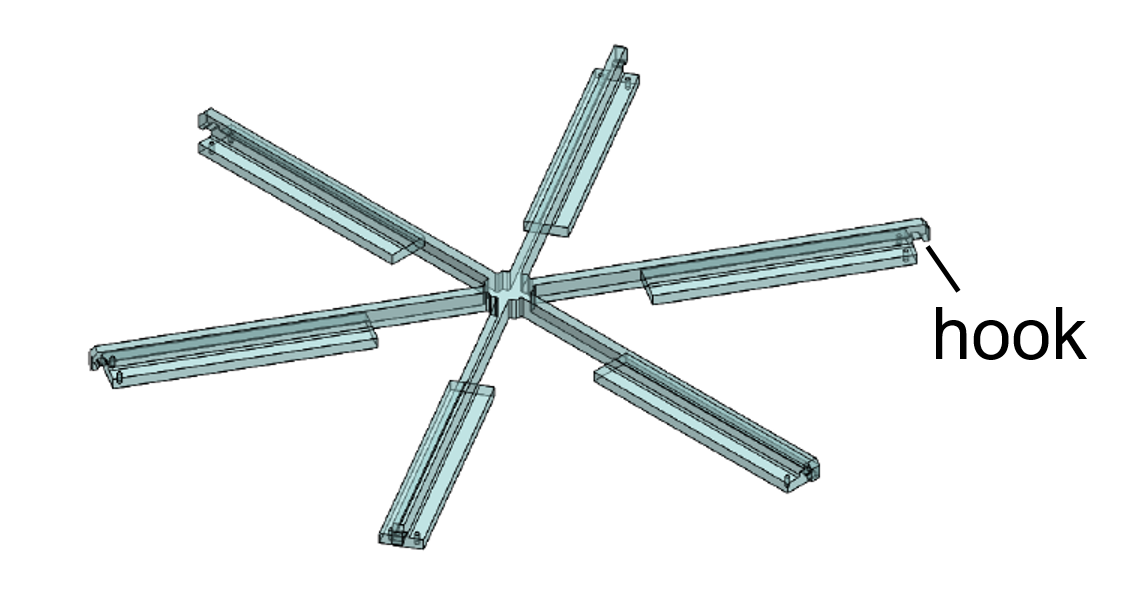}
			\caption{}
			\qquad
		\end{subfigure}
		\caption{(a) Support I-beam structure,  (b) Bracket}
		\label{fig:Ibeam}
	\end{figure}

	Figure \ref{fig:Acrylic_vessel} shows the design of the acrylic vessel. The main body consists of a cylindrical structure with a diameter and height of 3.6 m and 3.3 m, respectively, resulting in a capacity of 32 tonnes of DIN dissolved Gd-LS. Vessel optimization using stress simulations find the required thicknesses of 25 mm for the barrel and bottom, and 30 mm for the top-lid, respectively. Figure \ref{fig:Ibeam} show the design of the support I-beam and the bracket to support the  2,069 kg main acrylic cylinder structure. 
 The acrylic bracket is glued to the main cylinder. After the glue, an annealing process is performed to reduce internal stresses on all joint areas. The purpose of this annealing process is to prevent cracking of acrylic and to improve the mechanical and thermal properties. 
    The hooks of brackets are utilized to move the acrylic vessel with a crane. Each acrylic I-beam has a weight of $\sim$80 kg, and a length, largest width, and height of 120 cm, 20 cm, and 62.5 cm, respectively. The bottom-side of the six support I-beams are connected with bolts to the stainless steel tank, and I-beam top-side is connected to the bracket with bolts to fix the acrylic vessel in place. 
	
	A chimney on the top of the acrylic vessel will play a crucial role during liquid filling, liquid level monitoring, liquid level stabilizing, and detector calibration \cite{JSNS2_TDR}. The chimney is connected with 24 bolts and an o-ring to the main cylindrical part.
	
	\begin{figure}[htbp]
		\centering
		\includegraphics[width=.25\textwidth]{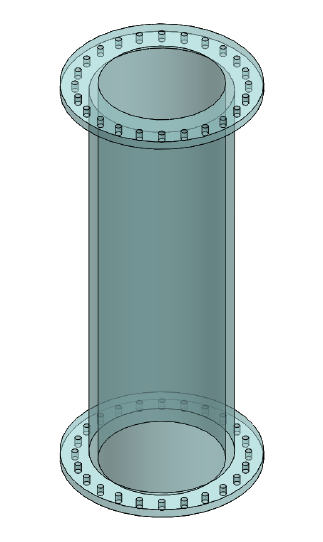}
		\caption{The acrylic vessel chimney structure with a diameter and height of 57.0 cm, and 150.7 cm, respectively.}
		\label{fig:trans}
	\end{figure}

	\section{Measured properties of the acrylic vessel}
	\subsection{Transmittance}

	Figure \ref{fig:trans} (blue) shows the transmittance of the PMMA acrylic 
 board with 25 mm thickness, the same thickness used for bottom and barrel of the real vessel. The optical properties of liquid 
 scintillator, including spectral emission profile and transmittance, and the quantum efficiency of a JSNS$^2$-II PMT (type R7081 made by Hamamatsu 
 Photonics~\cite{CITE:Hamamatsu}) are overlaid for direct comparison.
 The DIN dissolved Gd-LS emits optical photons with a  wavelength longer
 than 370 nm, and has a transmittance better than 90 \% from 420 nm. As a result, the 
 scintillation photons received at the PMTs are distributed between 400 nm $\sim$ 550 nm, with a 
 peak around 420 nm. The acrylic vessel provides excellent transmittance to the essential wavelength 
 region of the scintillation light and that of high PMT acceptance.

 	\begin{figure}[htbp]
		\centering
		\includegraphics[width=.62\textwidth]{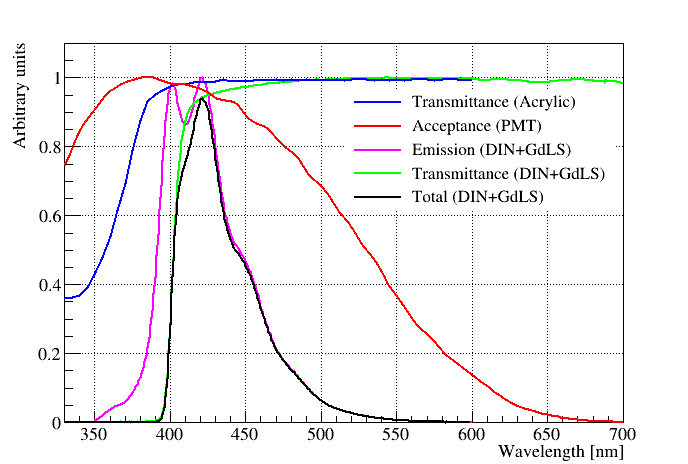}
		\caption{Optical properties of DIN dissolved Gd-LS and the acrylic vessel. }
		\label{fig:trans}
	\end{figure}
 

	\subsection{Thickness}
 	
\begin{figure}[htbp]
		\centering
		\begin{subfigure}[b]{0.325\textwidth}
			\includegraphics[width=1.\textwidth]{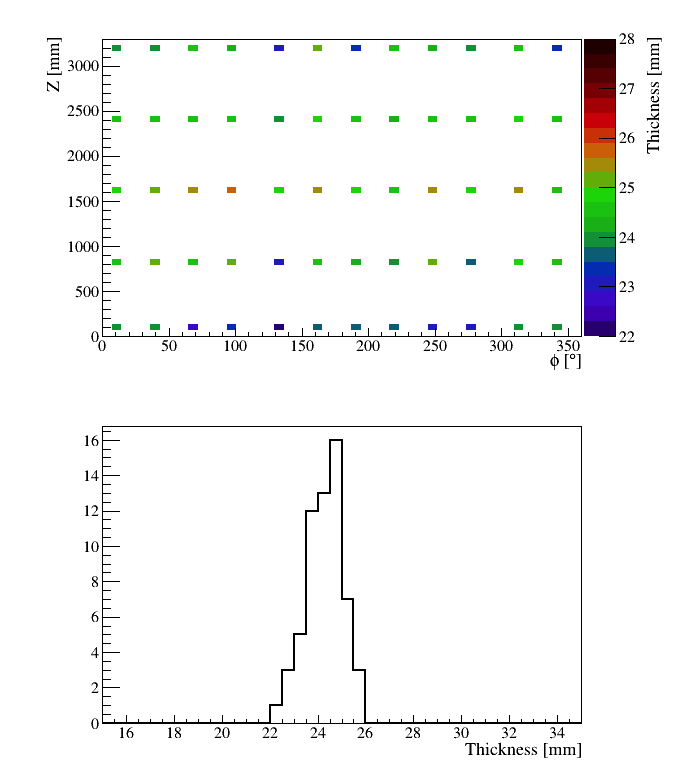}
			\caption{ barrel }
			\qquad
		\end{subfigure}
		\begin{subfigure}[b]{0.325\textwidth}
			\includegraphics[width=1.\textwidth]{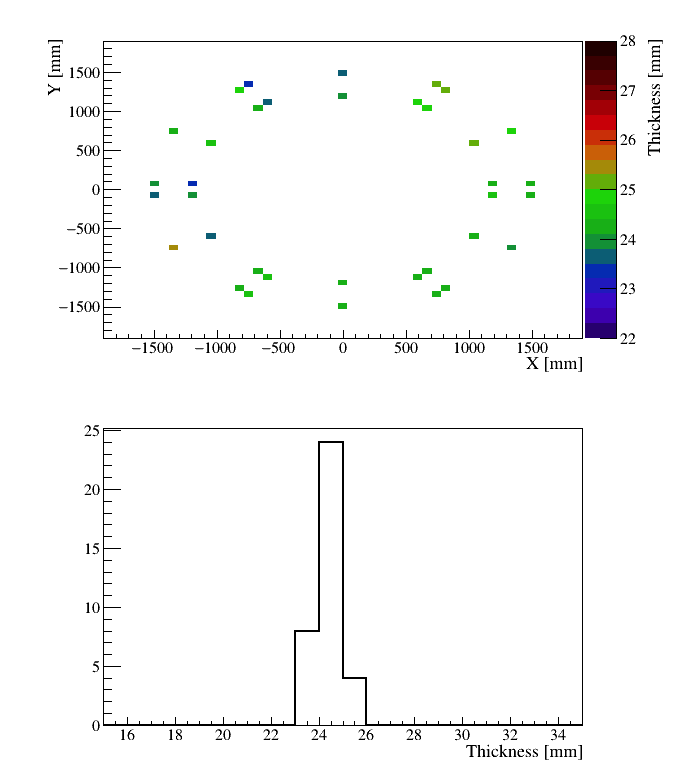}
			\caption{ bottom }
			\qquad
		\end{subfigure}
		\begin{subfigure}[b]{0.325\textwidth}
			\includegraphics[width=1.\textwidth]{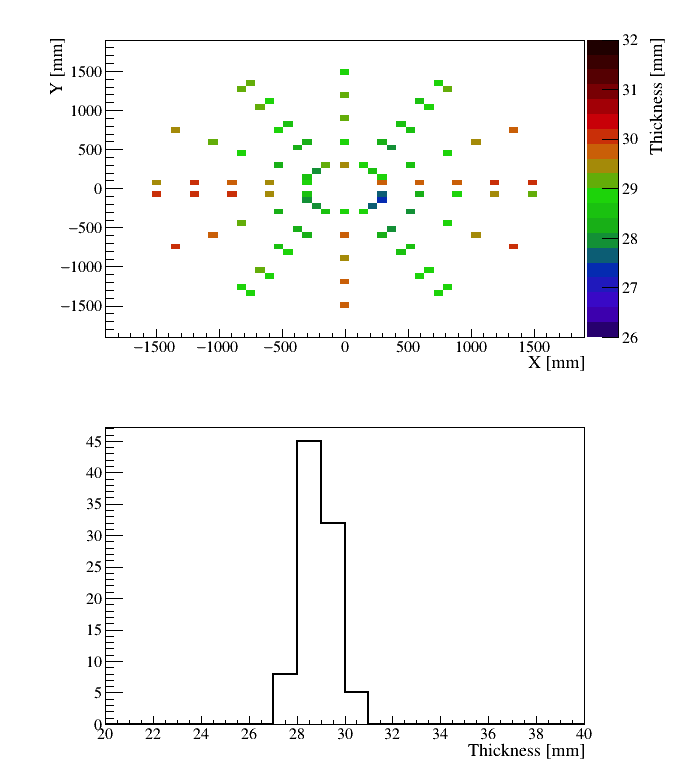}
			\caption{ top }
			\qquad
		\end{subfigure}
		\caption{The top panels show the measured acrylic thickness, represented by the colors, at a number of sample points. The distributions of measured values in each region are shown in the bottom panels.}
		\label{fig:Thick}
	\end{figure}

	The acrylic vessel thickness is measured and compared with design specifications using an ultrasonic thickness gauge at various sample positions of the barrel, bottom, and top regions. Figure \ref{fig:Thick} shows the positions and the results of thickness measurements. The average thickness and RMS are summarized in table \ref{Tab:thick}. These values are $\sim$1 mm less than designed values, and these differences were caused by the process of polishing the acrylic vessel surface. The measured thicknesses are acceptable for the detector design and will be considered in the detector simulation.

 \begin{table}[h]
	\begin{center}
		\begin{tabular*}{0.45\textwidth}{ c | c  | c | c }\hline\hline
			Region & Barrel & Bottom & Top \\\hline
			Mean & 24.3 mm & 24.3 mm & 28.9 mm\\\hline
   		RMS & 0.7 & 0.5 & 0.6\\
			\hline\hline
		\end{tabular*}
		\caption{Measured thickness of the acrylic vessel.}
		\label{Tab:thick}
	\end{center}
\end{table}

	\subsection{Leakage test by gas}
	Since the acrylic vessel will have to strictly separate the DIN dissolved Gd-LS in the target region from the pure LS in the catcher region, a leak test of the acrylic vessel prior to its installation and filling is essential.

	\begin{figure}[htbp]
		\centering
		\includegraphics[width=.7\textwidth]{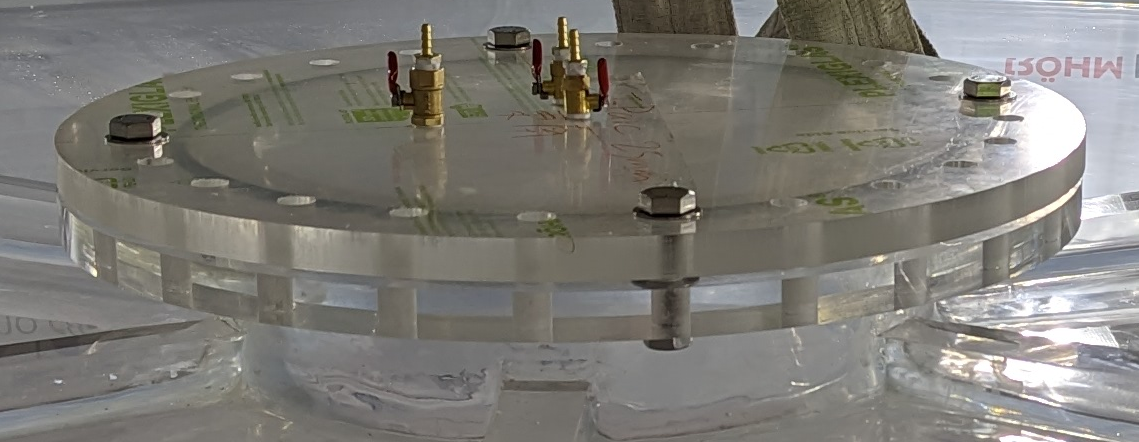}
		\caption{The picture around the special temporary flange made by an acrylic for the chimney hole. The valves are used for the gas injection.}
		\label{fig:Ccover}
	\end{figure}
	
	\begin{figure}[htbp]
		\centering
		\includegraphics[width=.9\textwidth]{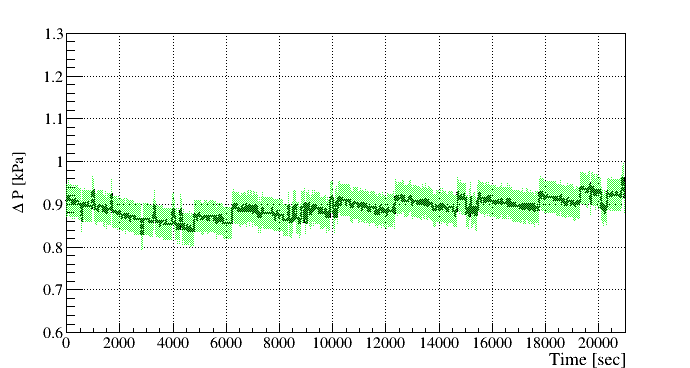}
		\caption{The result of the leak test using N$_{2}$ gas. The horizontal axis is the time after gas injection, and the vertical axis is the pressure difference. The uncertainty is shown as the green hatched area, which are due to the uncertainty of the instruments.}
		\label{fig:Leak}
	\end{figure}

	The leak test was performed using nitrogen (N$_{2}$) gas instead of liquid. The chimney opening was closed with the special temporary acrylic flange and an o-ring during the leak test as shown in figure \ref{fig:Ccover}. The pressure difference between the inside and outside of the acrylic vessel was measured using a digital differential pressure gauge (GC-62~\cite{CITE:GC62}). In addition, the temperature and the atmospheric pressure were measured by an ambient sensor-recorder (TR73-U~\cite{CITE:TR73-U}) simultaneously to correct these 
    parameters. The atmospheric pressure directly affects to the differential pressure, and the temperature also affects to the pressure linearly. For the latter, the correction was performed by the ideal gas equation of state.  
    For the leak test, the initial pressure difference after N$_{2}$ injection was set to be $\sim$0.9 kPa and monitored over a period of $\sim$6 hours. The differential pressure was stable as shown in figure \ref{fig:Leak}, and no leak was detected.

	\section{Summary}
	
	The acrylic vessel for the far detector of JSNS$^{2}$-II was constructed. The acrylic vessel is cleaned and washed with a 1\% of Alconox solution, followed by multiple rinses with ultrapure water. We confirmed that the acrylic vessel is in good agreement with its design specifications, including vessel thickness and optical property. It provides excellent transparency for the spectral wavelength range of the emission spectrum from the DIN dissolved Gd-LS. A leak test using N$_{2}$ gas, confirmed the sealing status of the acrylic vessel.

	\acknowledgments
We thank the J-PARC staff for their support. We acknowledge the support of the Ministry of Education, Culture, Sports, Science, and Technology (MEXT)
and the JSPS grants-in-aid: 16H06344, 16H03967 and
20H05624, Japan. 
This work is also supported by the National Research Foundation of Korea (NRF): 2016R1A5A1004684, 2017K1A3A7A09015973, 2017K1A3A7A09016426,
2019R1A2C3004955, 2016R1D1A3B02010606, 2017R1A2B4011200, 2018R1D1A1B07050425,
2020K1A3A7A09080133, 2020K1A3A7A09080114, 2020R1I1A3066835, 2021R1A2C1013661 and 2022R1A5A1030700. Our work has also been supported by a fund from the BK21 of the NRF. The University of Michigan gratefully acknowledges the support of the Heising-Simons Foundation. This work conducted at Brookhaven
National Laboratory was supported by the U.S. Department of Energy under Contract DE-AC02- 98CH10886. The work of the University of Sussex is supported by the Royal Society grant no. IESnR3n170385. We also thank the Daya Bay Collaboration for providing the Gd-LS, the RENO collaboration for providing the LS and PMTs, CIEMAT for providing the splitters, Drexel University for providing the FEE circuits and Tokyo Inst. Tech for providing FADC boards

	

\begin{thebibliography}{99}
		\bibitem{JSNS2_proposal}
		JSNS2 collaboration, Proposal: A Search for Sterile Neutrino at J-PARC Materials and Life Science Experimental Facility, arXiv:1310.1437.
		
		
		\bibitem{JSNS2_TDR}
		S. Ajimura et al., Technical Design Report (TDR): Searching for a Sterile Neutrino at J-PARC MLF (E56, JSNS$^2$), arXiv:1705.08629.
		
		\bibitem{cite:LSND} 
		LSND collaboration, Evidence for anti-muon-neutrino —> anti-electron-neutrino oscillations from the LSND experiment at LAMPF, \emph{Phys. Rev. Lett.} \textbf{77} (1996) 3082 [nucl-ex/9605003].
		
		\bibitem{JSNS2_II_TDR}
		S. Ajimura et al., Proposal: JSNS$^2$-II,
		arXiv:2012.10807

		\bibitem{DC_detector}
		Double Chooz collaboration, The Double Chooz antineutrino detectors, 
		\emph{Eur. Phys. J. C} {\bf 82} (2022) 804 [arXiv:2201.13285].
  
		\bibitem{RENO_acrylic}
		K.S. Park et al., Construction and properties of acrylic vessels in the RENO detector, 
		\emph{Nucl. Instrum. Meth. A} \textbf{686} (2012) 91.
		
		\bibitem{DayaBay_acrylic}
		H.R. Band et al., Acrylic Target Vessels for a High-Precision Measurement of theta13 with the Daya Bay Antineutrino Detectors,
		2012 \emph{JINST} 7 P06004 [arXiv:1202.2000].

  	\bibitem{CITE:Hamamatsu}
		Data sheet of PMT R7081 by Hamamatsu Photonics, Japan,         {\url{https://www.hamamatsu.com/content/dam/hamamatsu-photonics/sites/documents/99_SALES_LIBRARY/etd/LARGE_AREA_PMT_TPMH1376E.pdf}}

        \bibitem{CITE:GC62}
		{\url{http://products.naganokeiki.co.jp/assets/files/3035/E-GC62Q20130402.pdf}.}
  
        \bibitem{CITE:TR73-U}
		{\url{https://tandd.com/product/tr73u/}.}
		
	\end{thebibliography}
	

\end{document}